\documentstyle[12pt]{article}
\begin{document}
\title{A NOTE ON $3 + 1$ DIMENSIONALITY}
\author{B.G. Sidharth$^*$\\
Centre for Applicable Mathematics \& Computer Sciences\\
B.M. Birla Science Centre, Adarsh Nagar, Hyderabad - 500063 (India)}
\date{}
\maketitle
\footnotetext{E-mail:birlasc@hd1.vsnl.net.in;Fax:0091-40-237266;Tel:0091-40-235081}
\begin{abstract}
Recent work by Castro, Granik and El Naschie has given a rationale for the
three dimensionality of our physical space within the framework of cantorian
fractal space time using similar ideas of quantized fractal space time and
noncommutativity. We also deduce the same result. Interestingly this is also
seen to provide a rationale for an unproven conjecture of Poincare.
\end{abstract}
In a recent paper, Castro, Granik and El Naschie have given a rationale for
the three dimensionality of our physical space within the framework of a Cantorian
fractal space time and El Naschie's earlier work thereon\cite{r1,r2}. An ensemble
is used and the value for the average dimension involving the golden mean is
deduced close to the value of our $3 + 1$ dimensions. We now make a few remarks
based on an approach which is in the spirit of the above considerations.\\
Our starting point is the fact that the fractal dimension of a quantum path
is two, which, it has been argued is described by the coordinates $(x, ict)$\cite{r3}. Infact this leads to
discretized space time, and a non commutative geometry as in the approach
of El Naschie, Castro and Granik, and further the Dirac equation of the spin half
electron is seen to emerge from these considerations\cite{r4}. Given the
spin half, it is then possible to deduce the dimensionality of an ensemble
of such particles, which turns out to be three\cite{r5,r6}.\\
There is another way of looking at this. If we generalise from the one space
dimensional case and the complex $(x,t)$ plane to three dimensions, we infact
obtain the four dimensional case and the Theory of Quarternions, which are based
on the Pauli Spin Matrices\cite{r7}. As has been noted by Sachs, had Hamilton
identified the fourth coordinate in the above generalisation with time, then
he would have anticipated Special Relativity itself. It must be observed that
the Pauli Spin Matrices which denote the Quantum Mechanical spin half form,
again, a non commutative structure.\\
Curiously enough the above consideration in the complex plane can have an
interesting connection with an unproven nearly hundred year old conjecture of
Poincare. This was, the fact that closed loops could be shrunk to points as
happens on the two dimensional surface of a sphere is also valid in three
dimensions. Such closed loops defining a simply connected space are described
by closed contours in complex space which have an integral value for the
index\cite{r8}. For the surface of the sphere the axes are replaced by great
circles and the coordinates by the longitude and latitude. If we now generalise
to three dimensions, we end up instead with the above four dimensional
Quarternion theory of Hamilton - the three dimensional case is not possible.
We have instead three space dimensions, and a fourth dimension, which as
seen above can also be recovered from the doubly connected or spinorial
behaviour. Interestingly this also provides a rationale for Poincare's
conjecture not being provable.

\end{document}